\documentclass[10pt]{artikel3}
\usepackage{amsfonts,amsmath,amssymb}
\usepackage{float,braket}
\usepackage{pslatex}
\usepackage{subfigure}
\usepackage{graphicx,wrapfig}
\usepackage[parfill]{parskip}
\newcommand{\gf}{\ensuremath{\mathrm{gf}}}
\newcommand{\YM}{\ensuremath{\mathrm{YM}}}

\newcommand{\p}{\partial}

\newcommand{\occ}{\overline{c}}

\renewcommand{\d}{\ensuremath{\mathrm{d}}}
\newcommand{\e}{\ensuremath{\mathrm{e}}}
\usepackage[small,bf]{caption}
\newcommand{\sect}[1]{ \section{#1} \setcounter{equation}{0} }

\begin{document}
\title{{\bf A purely algebraic construction of a gauge and renormalization group invariant scalar glueball operator}}
\author{D.~Dudal\thanks{david.dudal@ugent.be}\,\,$^{a,b}$, S.~P.~Sorella\thanks{sorella@uerj.br\;;\;Work supported by FAPERJ, Funda{\c
c}{\~a}o de Amparo {\`a} Pesquisa do Estado do Rio de Janeiro, under
the program {\it Cientista do Nosso Estado},
E-26/100.615/2007.}\,\,$^c$,
N.~Vandersickel\thanks{nele.vandersickel@ugent.be}\,\,$^b$,
H.~Verschelde\thanks{henri.verschelde@ugent.be}\,\,$^b$\\\\
\small $^a$ \textnormal{Center for Theoretical Physics, Massachusetts Institute of Technology,} \\\small \textnormal{77 Massachusetts Avenue, Cambridge, MA 02139, USA}\\\\
\small $^b$ \textnormal{Ghent University, Department of Mathematical
Physics and Astronomy} \\ \small \textnormal{Krijgslaan 281-S9, 9000
Gent,
Belgium}\\\\
\small $^c$  \textnormal{Departamento de F\'{\i }sica Te\'{o}rica,
Instituto de
F\'{\i }sica, UERJ - Universidade do Estado do Rio de Janeiro}\\
\small   \textnormal{Rua S\~{a}o Francisco Xavier 524, 20550-013
Maracan\~{a}, Rio de Janeiro, Brasil }\normalsize}
\date{}
\maketitle

\begin{abstract}
\noindent This paper presents a complete algebraic proof of the
renormalizability of the gauge invariant $d=4$ operator
$F_{\mu\nu}^2(x)$ to all orders of perturbation theory in pure
Yang-Mills gauge theory, whereby working in the Landau gauge. This
renormalization is far from  being trivial as mixing occurs with
other $d=4$ gauge variant operators, which we identify explicitly.
We determine the mixing matrix $Z$ to all orders in perturbation
theory by using only algebraic arguments and consequently we can
uncover a renormalization group invariant by using the anomalous
dimension matrix $\Gamma$ derived from $Z$. We also present a future
plan for calculating the mass of the lightest scalar glueball with
the help of the framework we have set up.
\end{abstract} 
\vspace{-14cm} \hfill MIT-CTP 4004\vspace{14cm}


\sect{Introduction} Quantum Chromodynamics (QCD) is the theory of
the strong interactions which describes the force between quarks and
gluons. At high energies, this theory is asymptotically free, while
at low energies, only colorless bound states appear in nature due to
confinement. So far, the mechanism of confinement is still poorly
understood,  since at low energies non-perturbative aspects play an
important role. It is therefore of paramount importance to study
objects which can bring us closer to the understanding of
confinement. Glueballs are highly interesting candidates, as they
are entirely composed of gluons, and therefore the gauge field
itself is a crucial ingredient. For standard hadronic particles on
the other hand, also matter fields are indispensable. Hence,
glueballs have been widely investigated, experimentally, on the
lattice and in various theoretical models \cite{Mathieu:2008me}.

So far, there is no clear experimental evidence for the existence of
glueballs. If glueballs are observable particles, they would
strongly mix with other states containing quarks.  Due to this
feature, a clear observation of a glueball state turns out to be
rather difficult. However, there are already many indications for
the existence of glueballs, and the debate is currently ongoing.  It
is worth mentioning here that several experiments are actually
running and other ones are planned to start in the next future: $
\overline{\mathrm{P}}$ANDA \cite{Bettoni:2005ut} , BES III
\cite{Chanowitz:2006wf} and GlueX \cite{Carman:2005ps} to name only
a few. Glueballs might also play an important role in the quark
gluon plasma, a case that will be studied at e.g. the heavy ion
collision experiment ALICE at CERN \cite{Alessandro:2006yt}.

As no clear experimental data is yet available, the output of
theoretical models ought to be compared with lattice data. In
lattice gauge theories, there is no doubt about the existence of
glueballs,  although lattice calculations are still limited as they
cannot determine the decay channels of glueballs. In contrast with
possible experimental data, lattice calculations can however also
consider pure gauge theory. A consensus on the lowest lying scalar
glueball mass in the pure gauge gauge theory has already been
reached : $M_{0^{++}} \sim 1.6 $ GeV for SU(3)
\cite{Morningstar:1999rf,McNeile:2002en,Chen:2005mg,Teper:1998kw,Meyer:2008tr,West:1997sz}.

Many theoretical models have been investigated and compared with the
lattice data. An extensive recent overview is given in
\cite{Mathieu:2008me}. Historically, the first model to describe
glueballs is called the MIT bag model \cite{Jaffe:1975fd}. In this
model, gluons are placed in a bag and confined by a boundary
condition and a constant energy density $B$. This model, however, is
rather phenomenological in nature. Other phenomenological models
assume the gluons to have an effective mass
\cite{Cornwall:1981zr,Bernard:1981pg}, which can be used to compose
effective (potential) theories in which the masses of the different
glueballs are calculated
\cite{Cornwall:1982zn,Barnes:1981ac,Szczepaniak:1995cw,Kaidalov:1999yd}.

A more direct way to deal with glueballs is by identifying suitable
gauge invariant operators, which carry the correct quantum numbers
to create/annihilate particular glueball states, and then
calculating the corresponding correlators to get information on the
mass. In particular, this route is followed in the widely used QCD
sumrule approach \cite{Novikov:1979va,Narison:2008nj}. For example,
the operator relevant for the lightest scalar glueball is
$F^2(x)\equiv F_{\mu\nu}^2(x)$, hence the study of the correlator
$\Braket{F^2(x) F^2(y)}$. One takes into account perturbative as
well as non-perturbative contributions, which are associated with
condensates and instantons \cite{Novikov:1979va,Shuryak:1982dp}.
Also in the AdS/QCD approach, glueball (correlators) have been
investigated based on the assumption that there is an approximate
dual gravity description \cite{Brower:2000rp,Forkel:2007ru}.

In the light of such correlator studies, it would be interesting to
investigate the correlator $\Braket{F^2(x) F^2(y)}$ within the
Gribov-Zwanziger framework. The Gribov-Zwanziger action
\cite{Zwanziger:1989mf} was constructed in order to analytically
implement the restriction to the Gribov region $\Omega$, defined as
the set of field configurations fulfilling the Landau gauge
condition and for which the Faddeev-Popov operator, \begin{eqnarray}
\mathcal{M}^{ab} &=& -\partial_{\mu}\left( \p_{\mu} \delta^{ab} + g
f^{acb} A^c_{\mu} \right) \,, \end{eqnarray} is strictly positive,
namely
\begin{eqnarray} \Omega &\equiv &\{ A^a_{\mu}, \, \partial_{\mu}
A^a_{\mu}=0, \, \mathcal{M}^{ab}  >0\} \,. \end{eqnarray} The
boundary, $\partial \Omega$, of the region $\Omega$ is called the
(first) Gribov horizon. This restriction is necessary to avoid the
appearance of Gribov gauge copies in the Landau gauge
\cite{Gribov:1977wm}. Unfortunately, there are still a number of
Gribov copies remaining, but the Gribov-Zwanziger action is so far
the best approximation available. The Gribov-Zwanziger action is
originally constructed as a non-local action. However, with the
introduction of new fields, one can localize this action into the
following form,
\begin{eqnarray} \label{GZaction}
S_{0} &=&S_{\YM}+\int \d^{4}x\,\left( b^{a}\partial_\mu
A_\mu^{a}+\overline{c}^{a}\partial _{\mu } D_{\mu }^{ab}c^{b}\right)
+\int \d^{4}x\left( \overline{\varphi }_{\mu}^{ac}\partial _{\nu}
D_{\nu }^{ab}\varphi _{\mu}^{bc}-\overline{\omega
}_{\mu}^{ac}\partial _{\nu} D_{\nu }^{ab}\omega _{\mu}^{bc} -g
\partial _{\nu }\overline{\omega }_{\mu}^{ac} f^{abm}
D_{\nu }^{be}c^{e}\varphi _{\mu}^{mc} \right) \nonumber\\ && -\gamma
^{2}g\int\d^{4}x\left( f^{abc}A_{\mu }^{a}\varphi _{\mu
}^{bc}+f^{abc}A_{\mu}^{a}\overline{\varphi }_{\mu }^{bc} +
\frac{4}{g}\left(N^{2}-1\right) \gamma^{2} \right)\,,
\end{eqnarray} whereby $S_\YM$ denotes the Yang-Mills action, the
fields $\left( \overline{\varphi
}_{\mu}^{ac},\varphi_{\mu}^{ac}\right) $ are a pair of complex
conjugate bosonic fields and $\left( \overline{\omega
}_{\mu}^{ac},\omega_{\mu}^{ac}\right) $ are anticommuting fields
needed to localize the original non-local action. The parameter
$\gamma$ is fixed by a gap equation and implements the restriction
to the Gribov region. This $\gamma$\, introduces a mass scale into
the theory which can consequently give rise to a nonvanishing pole
in the glueball correlator. By taking into account the dynamics of
the fields $(\overline \varphi , \varphi , \overline \omega,
\omega)$, it became clear that a second mass scale $M^2$ emerges
quite naturally, and this $M^2$ can also enter the glueball
correlator expression. For more details on the Gribov-Zwanziger
action, its renormalization and the dynamics of its constituent
fields, we refer to \cite{Dudal:2007cw, Dudal:2008sp}.

Once the operator $F_{\mu\nu}^2$ is introduced, the issues of
renormalization and mixing complicate matters at the quantum level.
If one wants to investigate $F_{\mu\nu}^2$ in a renormalizable
setting, one should introduce this operator into the action by
coupling it to a source $q$, and then renormalize that action.
Therefore, before scrutinizing the more complicated Gribov-Zwanziger
case, it is instructive to first completely investigate
$F_{\mu\nu}^2$ within the usual Yang-Mills theory, quantized in the
Landau gauge. Moreover, as $F_{\mu\nu}^2$ is not a renormalization
group invariant, one could look for a renormalization group
invariant operator, containing $F_{\mu\nu}^2$, since renormalization
group invariance is beneficial when looking at physical quantities,
in casu the glueball mass. This renormalization group invariant will
turn out to coincide with the trace anomaly. However, we would like
to avoid a direct use of the trace anomaly, as the renormalization
of the trace anomaly itself, through that of the energy momentum
tensor, is rather difficult and sometimes tricky
\cite{Collins:1976vm}. Therefore, we shall focus on the direct
renormalization of $F_{\mu\nu}^2$ in the Landau gauge, which also
turns out to be far from trivial as a mixing with other (non gauge
invariant) operators occurs. In 1974, \cite{KlubergStern:1974rs}
described the first attempt towards the renormalization of gauge
invariant operators. In this paper, the renormalization of
$F_{\mu\nu}^2$ at zero momentum was investigated to the first loop
order, and the renormalization group invariant containing
$F_{\mu\nu}^2$ determined. However, the paper
\cite{KlubergStern:1974rs} focused  only on the integrated
operator\footnote{Or the operator at zero momentum.}
$\int\d^4xF_{\mu\nu}^2$ . In addition, a generalization to the more
complicated Gribov-Zwanziger case does not seem straightforward to
implement in the language of \cite{KlubergStern:1974rs}. Also, no
clear proof of the higher order renormalization of the nonintegrated
operator $F_{\mu\nu}^2(x)$ can be found. Notice that the passing
from the integrated to the nonintegrated operator is not trivial,
see \cite{Collins:1984xc}, \S 12.6, and references in
\cite{Gracey:2002rf}.

In \cite{Joglekar:1975nu,Deans:1978wn,Espriu:1983zz}, one has
elaborated on the structure of the mixing matrix for the more
general case of non-integrated gauge invariant operators, while in
\cite{Henneaux:1993jn} a simplified proof of the renormalization of
gauge invariant operators has been given from the perspective of the
BRST cohomology of Yang-Mills gauge theories. In the light of this
reference \cite{Henneaux:1993jn}, the last paper published on these
issues focusses in particular on some infrared subtleties
\cite{Collins:1994ee}. One can appreciate the intrinsic difficulties
arising when studying gauge invariant operators at the quantum level
by noticing that in \cite{Collins:1976yq}, results of
\cite{Joglekar:1975nu} were used, while one of the authors of
\cite{Collins:1976yq} quotes the same paper \cite{Joglekar:1975nu}
again in later years in \cite{Collins:1994ee}, mentioning that he
finds the proof of \cite{Joglekar:1975nu} ``very hard to
understand''.

Although the cohomological proof of \cite{Henneaux:1993jn} is of
full generality, the results are of an abstract nature. In an
oversimplifying nutshell, it was shown that each BRST invariant
operator at the quantum level with ghost number zero, can be written
as a strict gauge invariant operator\footnote{This means containing
only the field strength and covariant derivatives.} plus BRST exact
piece, modulo terms that vanish upon using the equation of motions.
For practical computations, it is however not sufficient to know
which type of operators occur in the renormalization process, but
also the explicit knowledge of all these operators is necessary.

For completeness, we also mention \cite{Tarrach:1981bi}, concerning
the renormalization of $F_{\mu\nu}^2$ in the background gauge
formalism, which would however be of little use when looking at the
Gribov-Zwanziger generalization.

Based on all the foregoing arguments, we have found it instructive
to present in this paper a clean analysis of the renormalization of
the operator $F_{\mu\nu}^2$ to all orders of perturbation theory.
The proof shall be given in the framework of algebraic
renormalization \cite{Piguet:1995er}, and we shall retrieve a
renormalization matrix, restricted by various Ward identities. Next
to suitable adaptations of the usual Landau gauge Ward identities,
we also identify a new powerful identity, relevant in the discussion
of the renormalization matrix, which form shall be in perfect
agreement with the argumentation given in
\cite{Collins:1984xc,Collins:1994ee}. We stress that our analysis is
purely algebraic, and in this sense differs from the argumentation
given in \cite{Collins:1984xc,Collins:1994ee}. Moreover, we shall
also be able to completely fix the mixing matrix to all orders and
this without calculating any loop diagram. Multiple checks will be
presented, which will confirm the results. For example, we shall
recover in an independent fashion well-known nonrenormalization
relations in the Landau gauge \cite{Piguet:1995er}, here stemming
from the renormalization analysis of $F_{\mu\nu}^2$.

In summary, in section II we shall give an overview of the 3
different classes of operators which can mix with $F_{\mu\nu}^2$
before going into the detailed algebraic renormalization of
$F_{\mu\nu}^2$. In section III, the mixing matrix will be determined
to all orders in perturbation theory and, armed with this result, we
shall be able to construct a renormalization group invariant in
section IV. We end this paper with a discussion in section V.

\sect{Renormalization of the Yang-Mills action with inclusion of the
operator $F_{\mu\nu}^2$}
\subsection{Introduction}
The most natural way to study the lightest scalar glueball is by
determining the correlator $\Braket{F^2(x) F^2(y)}$. This correlator
can be obtained by adding the operator $F_{\mu\nu}^2$ to the
ordinary Yang-Mills action by coupling it to a source $q(x)$.

Indeed, the action we start from reads,
\begin{eqnarray}\label{notrenorm}
\Sigma_{\mathrm{n.r.}} &=&\underbrace{\int \d^4 x \frac{1}{4}
F_{\mu\nu}^2}_{ S_{\YM}} + \underbrace{\int \d^4 x\,\left(
b^{a}\partial_\mu A_\mu^{a}+\overline{c}^{a}\partial _{\mu }
D_{\mu}^{ab}c^b \right)}_{S_{\gf}}  + \int \d^4 x \frac{q}{4}
F_{\mu\nu}^2\,,
\end{eqnarray}
whereby $S_\gf$ is the Landau gauge fixing part. For the benefit of
the reader, let us already mention that this action is BRST
invariant,
\begin{eqnarray}
s  \Sigma_{\mathrm{n.r.}} &=& 0\,,
\end{eqnarray}
with all the BRST transformations of the fields and the source given
by
\begin{align}\label{BRST}
sA_{\mu }^{a} &=-\left( D_{\mu }c\right) ^{a}\,, & sc^{a} &=\frac{1}{2}gf^{abc}c^{b}c^{c}\,,   \nonumber \\
s\overline{c}^{a} &=b^{a}\,,&   sb^{a}&=0\,, & s q & = 0\,,
\end{align}
and $s$ nilpotent,
\begin{eqnarray}
s^2 &=& 0\,.
\end{eqnarray}
In this fashion, the correlator is given by
\begin{eqnarray}
\left[{\frac{\delta}{\delta q(y)}  \frac{\delta}{\delta q(x)}}
Z^c\right]_{q=0} &=& \Braket{F^2(x) F^2(y)}\,,
\end{eqnarray}
with $Z^c$ the generator of connected Green functions. However, it
will turn out that the action \eqref{notrenorm} is not
renormalizable. Indeed, as the operator $F_{\mu\nu}^2$ has mass
dimension 4, it could mix with other operators of the same
dimension. The question arises which kind of extra operators we need
to consider.

\subsection{3 classes of operators\label{sectieB}}
In general, we can distinguish between 3 different classes of
dimension 4 operators. Firstly, the class $C_1$ contains all the
truly gauge invariant operators. These are the BRST closed but not
exact operators like $F_{\mu\nu}^2$. These are constructed from the
field strength $F_{\mu\nu}^a$ and the covariant derivative
$D_{\mu}^{ab}$. Secondly, the class $C_2$ consists of BRST exact
operators, e.g.~$s(\overline{c}^a \p_\mu A_\mu^a)$. The third class
$C_3$ contains operators which will vanish upon using the equations
of motion, e.g.~$A_\mu^a \frac{\delta S}{\delta A_\mu^a}$, with $S=S_\YM+S_\gf$.

Now, one can intuitively easily understand that these 3 different
classes will mix in a certain way
\cite{Collins:1984xc,Collins:1994ee}. Firstly, bare operators from
the class $C_2$ cannot receive contributions from gauge invariant
operators ($C_1$). Indeed, taking the matrix element of a bare BRST
exact operator from $C_2$ between physical states will give a
vanishing result, if there would be a renormalized gauge invariant
contribution from $C_1$ in its expansion, there would be a
nonvanishing contribution, clearly a contradiction. Secondly, as a
$C_3$ operator will vanish upon using the equations of motion, while
a $C_1$- and a $C_2$ operator in general do not, a $C_3$ operator
cannot receive corrections from the $C_1$ and/or $C_2$ class.

Thus, the mixing matrix will have an upper triangular form,
\begin{eqnarray}\label{upper} \left(
\begin{array}{c} \mathcal F_0 \\ \mathcal L_0 \\ \mathcal H_0
\end{array} \right) &= & \left( \begin{array}{ccc} Z_{\mathcal
F\mathcal F}& Z_{\mathcal F\mathcal L}  & Z_{\mathcal F \mathcal H}
\\ 0  &Z_{\mathcal L\mathcal L} & Z_{\mathcal L \mathcal H}   \\ 0 &
0& Z_{\mathcal H \mathcal H}          \end{array} \right) \left(
\begin{array}{c} \mathcal F \\ \mathcal L \\ \mathcal H \end{array}
\right)\,,
\end{eqnarray}
whereby $\mathcal F$, $\mathcal L$, $\mathcal H$ are operators
belonging,  respectively,  to the $C_1$, $C_2$ and $C_3$ class.

We shall however not use these observations, and only rely on a
formal algebraic analysis \cite{Piguet:1995er}. All constraints on
e.g. the mixing matrix should be encoded in the Ward identities.

For further use, let us elaborate a bit more on the equation of
motion like terms, using a scalar field for notational simplicity. A
term $\sim \frac{\delta S}{\delta\varphi}$ shall give rise to
contact terms when taking expectation values. Using partial path
integration, one finds
\begin{eqnarray}\label{s}
\Braket{\varphi(x_1)\varphi(x_2)\ldots\varphi(x_{n+1}) \frac{\delta
S}{\delta \varphi(y)}}&=&\int
[\d\phi]\varphi(x_1)\varphi(x_2)\ldots\varphi(x_{n+1}) \frac{\delta
S}{\delta \varphi(y)}\e^{-S}\nonumber\\&=&-\int
[\d\Phi]\varphi(x_1)\varphi(x_2)\ldots\varphi(x_{n+1})\frac{\delta}{\delta
\varphi(y)}\e^{-S}\nonumber\\
&=&\int[\d\Phi] \frac{\delta}{\delta \varphi(y)}
\left[\varphi(x_1)\varphi(x_2)\ldots\varphi(x_{n+1})\right] \e^{-S}
\nonumber\\&=&
\sum_{k=1}^{n+1}\delta(x_k-y)\braket{\varphi(x_1)\varphi(x_2)\ldots\varphi(x_{k-1})\varphi(x_{k+1})\ldots\varphi(x_{n+1})}\,.
\end{eqnarray}
We used the symbolic notation $ \int[\d \phi]$ for the integration
over all the present fields. Introducing the $Z$-factors for the
fields $\varphi$ , one also learns that $\varphi(y)\frac{\delta
S}{\delta \varphi(y)}$ does not need any renormalization factor, and
thus that it is \emph{finite} when introduced into
correlators\footnote{The implied limit $x_{n+1}\to y$ might seem
problematic due to the appearance of a $\delta(0)$ in the last term
of the r.h.s. of \eqref{s}.  However, $\delta(0)=0$ in dimensional
regularization.}. Moreover, if $x_k\neq y$, $k=1,\ldots,n$, the
l.h.s.~of \eqref{s} will vanish as the r.h.s.~does. On the other
hand, it is easily recognized from \eqref{s} that the integrated
operator $\int \d^4y \varphi(y)\frac{\delta S}{\delta\varphi(y)}$ is
nothing more than a counting operator when inserted into a
correlator, i.e.
\begin{eqnarray}\label{count}
\Braket{\varphi(x_1)\varphi(x_2)\ldots\varphi(x_n)\int\d^4y\varphi(y)
\frac{\delta S}{\delta
\varphi(y)}}&=&n\Braket{\varphi(x_1)\varphi(x_2)\ldots\varphi(x_n)}\,.
\end{eqnarray}

\subsection{The starting action}
We can now propose a more complete starting action than
\eqref{notrenorm}. Besides the gauge invariant operator
$F_{\mu\nu}^2$ belonging to the first class $C_1$, we also introduce
the BRST closed operator $s (\overline \p c  A)\equiv s(\p_\mu
\overline{c}^a  A_\mu^a) $, coupled to a new dimensionless source
$\eta$. As we want this new source to only enter the cohomological
trivial part of the action, we shall introduce a  BRST doublet
$(\lambda, \eta)$,
\begin{eqnarray}
s \eta &=& \lambda  \,,
\end{eqnarray}
and add the following term to the action \eqref{notrenorm},
\begin{eqnarray}
\int \d^4 x s ( \eta \overline c^a \p_\mu A_\mu^a) &=& \int \d^4 x
(\lambda \p_\mu \overline c^a A_\mu^a + \eta ( \p_\mu b^a A_\mu^a +
\p_\mu \overline c^a D^{ab}_{\mu} c^{b})  )\,.
\end{eqnarray}
The BRST doublet-structure is highly useful  in order construct the
most general invariant counterterm \cite{Piguet:1995er}.

Hence, the classical starting action is given by
\begin{eqnarray}\label{klassiek}
S_{\mathrm{cl}} &=& S_{\YM} + \int \d^{4}x\,\left( b^{a}\partial_\mu
A_\mu^{a}+\overline{c}^{a}\partial _{\mu } D_{\mu}^{ab}c^b \right) +
\int \d^4 x q \underbrace{ \frac{1} {4} F_{\mu\nu}^2}_{\mathcal F} +
\int \d^4 x \lambda \p_\mu \overline c^a A_\mu^a +\int \d^4 x \eta
\underbrace{ \left(  \p_\mu b^a A_\mu^a + \p_\mu\overline c^a
D_\mu^{ab} c^b \right)}_{\mathcal L}\,.\nonumber\\
\end{eqnarray}
Later in this paper, we shall also introduce the equation of motion
terms from class $C_3$. Notice that in principle, also
$s(\overline{c}^a\p_\mu A_\mu^a)$ is another independent $d=4$ BRST
exact operator which could play a role. It shall however turn out
that the renormalization analysis closes without this operator,
therefore we decided to immediately discard it.

We can now proceed with the study of this action, using the
formalism of algebraic renormalization \cite{Piguet:1995er}. A first
step is to introduce a term $S_{\mathrm{ext}}$,
\begin{eqnarray}
S_{\mathrm{ext}}&=&\int \d^{4}x\left( -K_{\mu }^{a}\left( D_{\mu
}c\right) ^{a}+\frac{1}{2}gL^{a}f^{abc}c^{b}c^{c}\right) \,,
\end{eqnarray}
needed to define the nonlinear BRST transformations of the gauge and
ghost fields. $K_{\mu }^{a}$ and $L^{a}$ are two new sources,
invariant under the BRST symmetry $s$. Therefore, the enlarged
action is given by
\begin{eqnarray} \label{startactie}
\Sigma &=& S_{\YM} + \int \d^{4}x\,\left( b^{a}\partial_\mu A_\mu^{a}+\overline{c}^{a}\partial _{\mu } D_{\mu}^{ab}c^b \right) + \int \d^{4}x \left( -K_{\mu }^{a}\left( D_{\mu }c\right) ^{a}+\frac{1}{2}gL^{a}f^{abc}c^{b}c^{c}\right) \nonumber\\
&&+  \int \d^4 x q  \frac{1} {4} F_{\mu\nu}^2  + \int \d^4 x \lambda
\p_\mu \overline c^a A_\mu^a +\int \d^4 x \eta \left(  \p_\mu b^a
A_\mu^a + \p_\mu\overline c^a D_\mu^{ab} c^b \right)\,,
\end{eqnarray}
and it will reduce itself to equation \eqref{klassiek}, once the
sources $K_\mu^a$ and $L^a$ are set to zero at the end. Likewise,
also $\lambda$ can be set to zero at that point.

A second step in the process of algebraic renormalization is to
determine all the Ward identities obeyed by the action
\eqref{startactie}, which we have summarized here:
\begin{itemize}
\item The Slavnov-Taylor idenitity:
\begin{eqnarray}\label{slavnov}
\mathcal{S}(\Sigma ) &=&\int \d^{4}x\left( \frac{\delta
\Sigma}{\delta K_{\mu }^{a}}\frac{\delta \Sigma }{\delta
A_{\mu}^{a}}+\frac{\delta \Sigma }{\delta L^{a}}\frac{\delta \Sigma
}{\delta c^{a}}+b^{a}\frac{\delta \Sigma}{\delta \overline{c}^{a}} +
\lambda \frac{\delta \Sigma }{\delta \eta} \right) = 0 \,.
\end{eqnarray}
\item The Landau gauge condition:
\begin{eqnarray}
\frac{\delta \Sigma}{\delta b^{a}}&=&\p_\mu A_\mu^a -\p_\mu(\eta
A_\mu^a)\,.
\end{eqnarray}
\item The modified antighost equation:
\begin{eqnarray}
\frac{\delta \Sigma}{\delta \overline c^{a}}+\p_\mu\frac{\delta
\Sigma}{\delta K_{\mu}^a}-(\p_\mu \eta)\frac{\delta \Sigma}{\delta
K_\mu^a }&=&0 \,.
\end{eqnarray}
\item The ghost Ward identity:
\begin{eqnarray}
\int \d^{4}x\left( \frac{\delta }{\delta
c^{a}}+gf^{abc}\left(\overline{c}^{b}\frac{\delta }{\delta b^{c}}
\right)\right)\Sigma &=& g\int \d^{4}xf^{abc}\left(
K_{\mu}^{b}A_{\mu }^{c}-L^{b}c^{c}\right)\,.
\end{eqnarray}
The term $\Delta _{\mathrm{cl}}^{a}$, being linear in the quantum
fields $A_{\mu }^{a}$, $c^{a}$, is a classical breaking.
\item The extra integrated Ward identity:
\begin{eqnarray}
\int \d^4 x\left( \frac{\delta \Sigma}{\delta \lambda} - \eta
\frac{\delta \Sigma}{\delta \lambda} + \overline c^a \frac{\delta
\Sigma}{\delta b^a} \right)&=& 0\,.
\end{eqnarray}
\end{itemize}
 Apart from some small adaptations, the first 5
symmetries are similar to the ones in the ordinary Yang-Mills
action. Moreover, we also find an extra Ward identity w.r.t.~the new
doublet $(\lambda, \eta)$. This last identity  will enable us to
take into account in a purely algebraic way the effects related to
the composite operators coupled to the sources $(\lambda, \eta)$.
We underline here that this is the power of the algebraic formalism:
by a well chosen set of sources to introduce the relevant operators,
one can hope to find additional Ward identities  which, in turn,
will constrain the theory at the quantum level, including the
characterization of the most general  counterterm. As such, a good
choice of sources can considerably simplify the renormalization
analysis.

\subsection{The counterterm}
When we turn to the quantum  level, we can use these symmetries to
characterize the most general allowed invariant counterterm
$\Sigma^{c}$. Following the algebraic renormalization procedure,
$\Sigma^{c}$ is an integrated local polynomial in the fields and
sources with dimension bounded by four, and with vanishing ghost
number. The previous, nonanomalous, Ward identities imply the
following constraints on $\Sigma^c$:
\begin{itemize}
\item The linearized Slavnov-Taylor identity:
\begin{equation}\label{ward1}
\mathcal{B}_{\Sigma }\Sigma ^{c}=0\,, \qquad \mathcal{B}_{\Sigma
}^{2} = 0\,,
\end{equation}
\begin{eqnarray}
\mathcal{B}_{\Sigma} &=&\int \d^{4}x\left( \frac{\delta
\Sigma}{\delta K_{\mu }^{a}}\frac{\delta }{\delta
A_{\mu}^{a}}+\frac{\delta \Sigma }{\delta A_{\mu
}^{a}}\frac{\delta}{\delta K_{\mu}^{a}}+\frac{\delta \Sigma }{\delta
L^{a}}\frac{\delta }{\delta c^{a}}+\frac{\delta\Sigma }{\delta
c^{a}}\frac{\delta }{\delta L^{a}}+b^{a}\frac{\delta }{\delta
\overline{c}^{a}} + \lambda \frac{\delta }{\delta \eta} \right)\,.
\end{eqnarray}
\item The Landau gauge condition:
\begin{eqnarray}
\frac{\delta \Sigma^c}{\delta b^{a}}&=&0 \,.
\end{eqnarray}
\item The modified antighost equation:
\begin{eqnarray}\label{modifiedantighost}
\frac{\delta \Sigma^c}{\delta \overline c^{a}}+\p_\mu\frac{\delta
\Sigma^c}{\delta K_{\mu}^a}-(\p_\mu \eta)\frac{\delta \Sigma}{\delta
K_\mu^a }&=&0 \,.
\end{eqnarray}
\item The ghost Ward identity:
\begin{eqnarray}
\int \d^{4}x\left( \frac{\delta }{\delta
c^{a}}+gf^{abc}\left(\overline{c}^{b}\frac{\delta }{\delta b^{c}}
\right)\right)\Sigma^c &=& 0 \,.
\end{eqnarray}
\normalsize
\item The extra integrated Ward identity:
\begin{eqnarray}\label{ward2}
\int \d^4 x\left( \frac{\delta \Sigma^c}{\delta \lambda} - \eta
\frac{\delta \Sigma^c}{\delta \lambda}  \right)&=& 0 \,.
\end{eqnarray}
\end{itemize}
To construct the most general counterterm, Tables \ref{tabel1} and
\ref{tabel2}, listing the dimension and ghost number of the various
fields and sources, are useful.
\begin{table}[t]
\begin{center}
        \begin{tabular}{|c|c|c|c|c|}
        \hline
        & $A_{\mu }^{a}$ & $c^{a}$ & $\overline{c}^{a}$ & $b^{a}$  \\
        \hline
        \hline
        \textrm{dimension} & $1$ & $0$ &$2$ & $2$  \\
        \hline
        $\mathrm{ghost\, number}$ & $0$ & $1$ & $-1$ & $0$   \\
        \hline
        \end{tabular}
        \caption{Quantum numbers of the fields.}\label{tabel1}
      \end{center}  \end{table}
        \begin{table}[t]
        \begin{center}
    \begin{tabular}{|c|c|c|c|c|c|}
        \hline
        &$K_{\mu }^{a}$&$L^{a}$& $q$& $\eta$ & $\lambda$  \\
        \hline
        \hline
        \textrm{dimension} & $3$ & $4$ & $0$ & $0$ & 0    \\
        \hline
        $\mathrm{ghost\, number}$ & $-1$ & $-2$ & 0 & 0 & 1  \\
        \hline
        \end{tabular}
        \caption{Quantum numbers of the sources.}\label{tabel2}
\end{center}
\end{table}
There is however one subtlety concerning counterterms quadratic (or
higher) in the sources. Only looking at the dimensionality, the
ghost number and the constraints on the counterterm, it is a priori
not forbidden to consider terms of the the form $(q^2 \ldots)$,
$(\eta^2 \ldots)$, $(q \eta \ldots)$, $(q^3 \ldots)$, etc., i.e.
terms of quadratic and higher order in the sources. If these terms
are allowed, an infinite tower of counterterms would be generated
and it would be impossible to prove the renormalizability of the
action as new divergences are being generated, which cannot be
absorbed in terms already present in the starting action. However,
we can give a simple argument why we may omit this class of terms.
Assume that we would also introduce the following term of order
$q^2$ in the action,
\begin{eqnarray}\label{term}
&\sim&  \int \d^4 x q^2 \frac{F_{\mu\nu}^2}{4}\,.
\end{eqnarray}
Subsequently, when calculating the correlator, this term would give
rise to an extra contact term contribution,
\begin{eqnarray}\label{argument}
\left[\frac{\delta }{\delta q(z)}\frac{\delta }{\delta q(y)}\int [\d
\phi] \e^{- S} \right]_{q=0}  &=& \underbrace{\Braket{F^{2}(z)
F^{2}(y) }}_{\mathrm{term\ due\ to\ part\ in\ }q} +
\underbrace{\delta(y-z) \Braket{ F^2(y)}}_{\mathrm{term\ due\ to\
part\ in\ }q^2} \,.
\end{eqnarray}
As eventually we are only interested in the correlator for $z \not =
y$, we can thus neglect the term \eqref{term}. In fact, when looking
at the case $z =y$, we should also couple a source to the novel
composite operator $F^4$, which is not our current interest. We can
repeat this kind of argument for all other terms of higher order in
the sources.

There is one exception to the previous remark: we cannot neglect
higher order terms of  the type $(K q \ldots)$ and $(K \eta \ldots)$
due to the modified antighost equation,
 \begin{eqnarray}\label{modifiedantighost}
\frac{\delta \Sigma^c}{\delta \overline c^{a}}+\p_\mu\frac{\delta
\Sigma^c}{\delta K_{\mu}^a}-(\p_\mu \eta)\frac{\delta \Sigma}{\delta
K_\mu^a }&=&0 \,.
\end{eqnarray}
The second term of this equation  differentiates the counterterm
w.r.t.~the source $K^a_\mu$, while the first term w.r.t. the field
$\overline c^a$. Therefore, for the construction of the counterterm
fulfilling all the constraints, we still need to include terms of
order $K q$ and $K \eta$, as when deriving w.r.t~$K_\mu^a$, these
terms will become of first order in the sources, just as the term
$\propto \frac{\delta \Sigma^c}{\delta \overline c^a}$. However,  at
the end, after having completely characterized the counterterm, we
can ignore this class of terms again.

We are now ready to construct the counterterm. Firstly,
 making use of general results on the cohomology of
gauge theories \cite{Piguet:1995er}, the most general integrated
polynomial of dimension 4 in the fields and sources, with vanishing
ghost number and  which takes into account the previous remarks on
the  terms quadratic in the sources, can be written as
\begin{eqnarray}
\Sigma^c &=& a_{0}  \int \d^4 x\frac{1}{4}F_{\mu\nu}^2 +b_{0}  \int \d^4 x\frac{q}{4}F_{\mu\nu}^2+ \mathcal{B}_\Sigma \int \d^{4}\!x\, \biggl\{ a_{1}(K_{\mu}^{a}+\partial _{\mu} \overline{c}^{a})A_{\mu}^{a} + a_{2}\,L^{a}c^{a} + a_3 b^a\occ^a + a_4 gf^{abc}\occ^a\occ^b c^c  \biggr\} \nonumber\\
&& +  \mathcal{B}_\Sigma \int \d^{4}\!x\, \biggl\{ b_{1}q (K_{\mu}^{a}+\partial _{\mu} \overline{c}^{a})A_{\mu }^{a} +  b_{2}q \overline{c}^{a} \partial _{\mu} A_{\mu}^{a} + b_3 q b^a\occ^a + b_4 q gf^{abc}\occ^a\occ^b c^c  \biggr\} \nonumber\\
&&+ \mathcal{B}_\Sigma \int \d^{4}\!x\, \biggl\{c_{1} \eta
K_{\mu}^{a}A_{\mu }^{a} +c_{2}\eta\partial _{\mu}
\overline{c}^{a}A_{\mu }^{a}+ c_{3}\eta \overline{c}^{a} \partial _{\mu} A_{\mu}^{a}  +  c_4 \eta b^a\occ^a + c_5 \eta gf^{abc}\occ^a\occ^b c^c   \biggr\}\nonumber\\
&& + \mathcal{B}_\Sigma \int \d^{4}\!x\, \biggl\{d_1 \lambda
\overline c^a \overline c^a \biggr\} \,.
\end{eqnarray}
Secondly, we can simplify this counterterm by imposing all the
constraints \eqref{ward1}-\eqref{ward2}. After a certain amount of
algebra, we eventually obtain
\begin{eqnarray}
\Sigma^c&=& a_{0}  \int \d^4 x\frac{1}{4}F_{\mu\nu}^2 +  b_0 \int \d^4 x\frac{q}{4}F_{\mu\nu}^2  + a_{1}\int \d^{4}x\Biggl( A_{\mu}^{a}\frac{\delta S_{\YM}}{\delta A_{\mu }^{a}}+  A_{\mu}^{a}\frac{\delta \widehat{S}_{\YM}}{\delta A_{\mu }^{a}}  +K_{\mu }^{a}\partial _{\mu }c^{a} + \p_\mu \overline{c}^a \p_\mu c^a - \eta \p_\mu \overline{c}^a \p_\mu c^a  \Biggr) \nonumber\\
&& + b_{1} \int \d^{4}x\, q \Biggl( A_{\mu}^{a}\frac{\delta
S_{\YM}}{\delta A_{\mu }^{a}}  + K_{\mu }^{a}\partial _{\mu }c^{a} +
\p_\mu \overline{c}^a \p_\mu c^a \Biggr)\,,
\end{eqnarray}
with
\begin{equation}\label{exra}
    \widehat{S}_{\YM}=\frac{1}{4}\int \d^4x qF_{\mu\nu}^2 \,.
\end{equation}
Now that we have constructed the most general counterterm obeying
all the Ward identities, we can neglect, as previously described,
the term in $K q$. Therefore, the final counterterm becomes,
\begin{eqnarray}
\Sigma^c&=& a_{0}  \int \d^4 x\frac{1}{4}F_{\mu\nu}^2 +  b_0 \int \d^4 x\frac{q}{4}F_{\mu\nu}^2  + a_{1}\int \d^{4}x\Biggl( A_{\mu}^{a}\frac{\delta S_{\YM}}{\delta A_{\mu }^{a}}+  A_{\mu}^{a}\frac{\delta \widehat{S}_{\YM}}{\delta A_{\mu }^{a}}  +K_{\mu }^{a}\partial _{\mu }c^{a} + \p_\mu \overline{c}^a \p_\mu c^a - \eta \p_\mu \overline{c}^a \p_\mu c^a  \Biggr) \nonumber\\
&& + b_{1} \int \d^{4}x\, q \Biggl( A_{\mu}^{a}\frac{\delta
S_{\YM}}{\delta A_{\mu }^{a}}  + \p_\mu \overline{c}^a \p_\mu c^a
\Biggr)\,.
\end{eqnarray}

\subsection{Introducing the equations of motion}
We still have to introduce the equations of motion as described in
Section \ref{sectieB}, as these can enter the operator $\mathcal F$.
So far, we have found an action $\Sigma$ with corresponding
counterterm $\Sigma^c$. Let us perform the linear shift on the gluon
field $A^a_\mu$,
\begin{eqnarray}
A^a_\mu \rightarrow A^a_\mu + J A^a_\mu\,,
\end{eqnarray}
with $J(x)$ a novel local source. This way of introducing the
relevant gluon equation of motion operator shall turn out to be very
efficient, as it allows us to uncover the finiteness of this kind of
operator. Indeed, this shift basically corresponds to a redefinition
of the gluon field, and has to be consistently done in the starting
action and counterterm. Performing the shift in the action gives
rise to the following shifted action $\Sigma'$,
\begin{eqnarray}\label{eindactie}
 \Sigma'&=& S_{\mathrm{\YM}} + \int \d^{4}x\,\left( b^{a}\partial_\mu A_\mu^{a}+\overline{c}^{a}\partial _{\mu } D_{\mu}^{ab}c^b \right) +\int \d^{4}x \left( -K_{\mu }^{a}\left( D_{\mu }c\right) ^{a}+\frac{1}{2}gL^{a}f^{abc}c^{b}c^{c}\right) \nonumber\\
&&+  \int \d^4 x q  \frac{1} {4} F_{\mu\nu}^2  + \int \d^4 x \lambda \p_\mu \overline c^a A_\mu^a +\int \d^4 x \eta\left(  \p_\mu b^a  A_\mu^a + \p_\mu\overline c^a D_\mu^{ab} c^b \right)\nonumber\\
&&+  \int \d^4 x J  \underbrace{ A_\mu^a\frac{\delta S_{\YM}}{\delta
A_\mu^a}}_{\mathcal H}  +  \int \d^4 x  J  \left\{  - \p_\mu b^a
A_\mu^a  + g f_{akb} A_\mu^k c^b \p_\mu \overline c^a  \right\} \,,
\end{eqnarray}
where we see the relevant gluon equation of motion term, $\mathcal
H$, emerging. Again, we have neglected higher order terms in the
sources, as the argument \eqref{argument} still holds. Analogously,
we find a shifted counterterm,
\begin{eqnarray}
\Sigma'^c&=& a_{0}  \int \d^4 x\frac{1}{4}F_{\mu\nu}^2 +  b_0 \int \d^4 x\frac{q}{4}F_{\mu\nu}^2 + a_{1}\int \d^{4}x\Biggl( A_{\mu}^{a}\frac{\delta S_{\YM}}{\delta A_{\mu }^{a}}+  A_{\mu}^{a}\frac{\delta \widehat{S}_{\YM}}{\delta A_{\mu }^{a}}  +K_{\mu }^{a}\partial _{\mu }c^{a} + \p_\mu \overline{c}^a \p_\mu c^a - \eta \p_\mu \overline{c}^a \p_\mu c^a  \Biggr) \nonumber\\
&& + b_{1} \int \d^{4}x\, q \Biggl( A_{\mu}^{a}\frac{\delta
S_{\YM}}{\delta A_{\mu }^{a}}  + \p_\mu \overline{c}^a \p_\mu c^a
\Biggr)+  a_{0}\int \d^{4}x\Biggl( J A_{\mu}^{a}\frac{\delta
S_{\YM}}{\delta A_{\mu }^{a}} \Biggr)\nonumber\\&& + a_1 \int \d^4 x
J \left( 2 A_\mu^a \p_\mu \p_\nu A_\nu^a - 2 A_\mu^a \p^2 A_\mu^a +
9 g f_{abc} A_\mu^a A_\nu^b \p_\mu A_\nu^c + 4 g^2 f_{abc} f_{cde}
A_\mu^a A_\nu^b A_\mu^d A_\nu^e\right)\,,
\end{eqnarray}
where one can neglect again the higher order terms in the sources.

One could also introduce the other similar equation of motion terms,
by introducing linear shifts for the $b^a$, $c^a$, $\overline c^a$
fields. However, the corresponding equation of motion operators will
not mix with $F_{\mu\nu}^2$ and are therefore unnecessary to
establish the renormalizability of the action \eqref{eindactie}.

\subsection{Stability and the renormalization (mixing) matrix}
Finally, it remains to discuss the stability of the classical
action, i.e.~to check whether $\Sigma'^c$ can be reabsorbed in the
classical action $\Sigma' $ by means of a multiplicative
renormalization of the coupling constant $g$, the fields $\left\{
\phi =A,c,\overline{c},b\right\} $ and the sources $\left\{ \Phi =
L,K, q, \eta, \lambda, J \right\} $, namely
\begin{equation}
\Sigma' (g,\phi ,\Phi )+h \Sigma'^c= \Sigma (g_{0},\phi
_{0},\Phi_{0})+{\cal O}(h^{2})\,,  \label{stab}
\end{equation}
with $h$ the infinitesimal perturbation parameter. The bare fields,
sources and parameters are defined as
\begin{align}
 K _{0\mu}^{a}&~=~Z_{K }K _{\mu }^{a}\,,                                  & A_{0\mu }^{a} &=Z_{A}^{1/2}A_{\mu }^{a}\,,                                  &  g_{0}~&=~Z_{g}g\,, \nonumber \\
 L_{0}^{a}~&=~Z_{L}L^{a}\,,                                               & c_{0}^{a} &=Z_{c}^{1/2}c^{a} \,,  \nonumber \\
 q_0 ~&=~Z_{q }q\,,                                                       & \overline{c}_{0}^{a} &=Z_{\overline{c}}^{1/2}\overline{c}^{a}\,,   &    \nonumber \\
 \eta_{0}~&=~Z_{\eta}\eta\,,                                      & b_{0}^{a} &=Z_{b}^{1/2}b^{a}\,,   \nonumber\\
 J_{0}~&=~Z_{J}J\,,   \nonumber\\
\lambda_0 &= Z_\lambda \lambda\,.
\end{align}
We also propose the following mixing matrix,
\begin{equation}\label{zmatrixbis}
 \left(
  \begin{array}{c}
    q_0 \\
    \eta_0 \\
    J_0
  \end{array}
\right)=\left(
          \begin{array}{ccc}
            Z_{q      q} & Z_{q      \eta}  & Z_{q      J} \\
            Z_{\eta q} & Z_{\eta \eta}  & Z_{\eta J} \\
            Z_{J      q} & Z_{J      \eta}  & Z_{J      J}
          \end{array}
        \right)
\left(
  \begin{array}{c}
    q \\
    \eta \\
    J
  \end{array}
\right)\,,
\end{equation}
which will represent the mixing of the operators $\mathcal F$,
$\mathcal L$ and $\mathcal H$. If we try to absorb the counterterm
into the original action, we ultimately find,
\begin{eqnarray}\label{Z1}
Z_{g} &=&1-h \frac{a_0}{2}\,,  \nonumber \\
Z_{A}^{1/2} &=&1+h \left( \frac{a_0}{2}+a_{1}\right) \,,
\end{eqnarray}
and
\begin{eqnarray}\label{Z2}
Z_{\overline{c}}^{1/2} &=& Z_{c}^{1/2} = Z_A^{-1/4} Z_g^{-1/2} = 1-h \frac{a_{1}}{2}\,, \nonumber \\
Z_{b}&=&Z_{A}^{-1}\,, \nonumber\\
Z_{K }&=&Z_{c}^{1/2}\,,  \nonumber\\
Z_{L} &=&Z_{A}^{1/2}\,,
\end{eqnarray}
results which are known from the renormalization of the original
Yang-Mills action in the Landau gauge \cite{Piguet:1995er}.

In addition, we also find the following mixing matrix
\begin{eqnarray}\label{mixingmatrix}
\left(
          \begin{array}{ccc}
            Z_{q      q} & Z_{q      \eta}  & Z_{q      J} \\
            Z_{\eta q} & Z_{\eta \eta}  & Z_{\eta J} \\
            Z_{J      q} & Z_{J      \eta}  & Z_{J      J}
          \end{array}
        \right) &=& \left(
          \begin{array}{ccc}
            1 + h (b_0 - a_0) & 0  & 0 \\
            h b_1 & 1  & 0 \\
            h b_1 & 0  & 1
          \end{array}
        \right)\,,
\end{eqnarray}
and for completeness, the $Z$-factor of $\lambda$ reads,
\begin{eqnarray}
Z_{\lambda} &=& Z_{c}^{-1/2} Z_A^{-1/2} \,,
\end{eqnarray}
as the counterterm does not contain the source $\lambda$.

Once having this mixing matrix at our disposal, we can of course
pass to the corresponding bare operators. For this, we shall need
the inverse of the mixing matrix \eqref{mixingmatrix},
\begin{equation}\label{zmatrixbis}
 \left(
  \begin{array}{c}
    q \\
    \eta \\
    J
  \end{array}
\right)=\left(
          \begin{array}{ccc}
            \frac{1}{Z_{q    q}} & 0 & 0 \\
            -\frac{Z_{Jq}}{ Z_{q    q}} & 1 & 0 \\
            -\frac{Z_{Jq}}{ Z_{q    q}} & 0  & 1
          \end{array}
        \right)
\left(
  \begin{array}{c}
    q_0 \\
    \eta_0 \\
    J_0
  \end{array}
\right)\,.
\end{equation}
Now we can determine the corresponding mixing matrix for the
operators, since insertions of an operator correspond to derivatives
w.r.t. to the appropriate source of the generating functional
$Z^c(q,\eta,J)$. In particular,
\begin{eqnarray}\label{ren1}
    \mathcal{F}_0 &=& \frac{\delta Z^c(q, \eta, J)}{\delta q_0} = \frac{\delta q}{\delta q_0}\frac{\delta Z^c(q, \eta, J) }{\delta q}+\frac{\delta \eta}{\delta q_0}\frac{\delta Z^c(q, \eta, J)}{\delta \eta } + \frac{\delta J}{\delta q_0}\frac{\delta Z^c(q, \eta, J)}{\delta J } \nonumber\\
    \Rightarrow  \mathcal{F}_0 &=&\frac{1}{Z_{q    q}}\mathcal{F}-\frac{Z_{Jq}}{ Z_{q    q}} \mathcal{G}  -\frac{Z_{Jq}}{ Z_{q    q}} \mathcal
    H\,,
\end{eqnarray}
and similarly for $\mathcal{G}_0$ and $\mathcal H_0$. In summary, we
find
\begin{eqnarray}\label{operatormatrix}
 \left(
  \begin{array}{c}
    \mathcal F_0 \\
    \mathcal L_0 \\
    \mathcal H_0
  \end{array}
\right) &= & \left(
          \begin{array}{ccc}
            Z_{qq}^{-1}& -Z_{Jq}Z_{qq}^{-1}  &-Z_{Jq}Z_{qq}^{-1} \\
            0  &1 & 0   \\
           0 & 0& 1          \end{array}
        \right)
        \left(
\begin{array}{c}
    \mathcal F \\
    \mathcal L \\
    \mathcal H
  \end{array}
\right)\,.
\end{eqnarray}
From this matrix we can make several interesting observations.
Firstly, we see that the operator $\frac{1}{4}F_{\mu\nu}^2 ~(=
\mathcal F)$ indeed required the presence of the BRST exact operator
$\mathcal L$ and  of the gluon equation of motion operator $\mathcal
H$ as these operators are ``hidden'' in the bare operator $\mathcal
F$. Secondly, we do retrieve an upper triangular matrix, in
agreement with the earlier description in \eqref{upper}. Moreover,
we also find that the BRST exact operator $\mathcal L$ does not mix
with $\mathcal H$, a mixing which is in principle allowed, but has a
$Z$-factor equal to $1$. This can be nicely understood: the
integrated BRST exact operator is in fact proportional to a sum of
two (integrated) equations of motion terms,
\begin{eqnarray}\label{countcon}
\int \d^4 x ( \p_\mu b^a A_\mu^a + \p_\mu \overline c^a D_\mu^{ab}
c^{b}) &=& - \int \d^4 x ( b^a \p_\mu A_\mu^a +
 \occ^a \p_\mu D_\mu^{ab} c^{b})=-\int \d^4 x\left(b^a\frac{\delta S}{\delta b^a}+\occ^a\frac{\delta S}{\delta \occ^a}\right)   \,,
\end{eqnarray}
and therefore it does  not mix with other operators, just like
$\mathcal H$.

\sect{The mixing matrix to all orders} In this section, we shall
demonstrate that we can determine the mixing matrix
\eqref{operatormatrix} exactly, i.e.  to all orders of perturbation
theory. For this purpose, we shall follow the lines of
\cite{Brown:1979pq}, suitably adapted to the gauge theory under
study. We start with the following most general $(n + 2m + r)$-point
function defined as,
\begin{eqnarray}\label{definitie}
\mathcal G^{n + 2m + r}(x_1, \ldots, x_n, y_1, \ldots, y_n,\hat y_1,\ldots, \hat y_n, z_1, \ldots, z_n ) &=& \Braket{A(x_1)\ldots A(x_n) c(y_1)\ldots c(y_m ) \overline c(\hat y_1) \overline c (\hat y_m) b(z_1)\ldots b(z_r)  }\nonumber\\
&&\hspace{-2.3cm}= \int [\d \phi] A(x_1)\ldots A(x_n) c(y_1)\ldots
c(y_m ) \overline c(\hat y_1) \overline c (\hat y_m) b(z_1)\ldots
b(z_r) \e^{- S} \,,
\end{eqnarray}
with the action $S$ given by
\begin{eqnarray}
S &=& S_{\YM} + S_{\gf}\,
\end{eqnarray}
We have immediately assumed that there is an equal amount of ghost
and antighost fields present as in any other case, the Green
function \eqref{definitie} would be zero, due to ghost number
symmetry. Subsequently, from the definition \eqref{definitie}, we
can immediately write down the connection between the renormalized
Green function and the bare Green function,
\begin{eqnarray}\label{connectie}
\mathcal G^{n + 2m + r} = Z_A^{-n/2} Z_c^{-m}Z_b^{-r/2} \mathcal
G_0^{n + 2m + r}\,.
\end{eqnarray}
From the previous equation, we shall be able to fix all the matrix
elements of expression \eqref{operatormatrix}, based on the
knowledge that
\begin{eqnarray}
\frac{\d \mathcal G^{n + 2m + r} }{\d g^2}
\end{eqnarray}
is finite.

We start by applying the chain rule when deriving the right hand
side of equation \eqref{connectie} w.r.t.~$g^2$. We find,
\begin{eqnarray}\label{finalgoal2}
\frac{\p \mathcal G^{n'} }{\p g^2} &=& \frac{\p g_0^2 }{ \p g^2}
\frac{\p \mathcal G_0^{n'} }{\p g_0^2} Z_A^{-n/2} Z_c^{-m}Z_b^{-r/2}
+ \frac{\p  Z_A^{-n/2} }{ \p g^2}  Z_c^{-m}Z_b^{-r/2}    \mathcal
G_0^{n'} +    Z_A^{-n/2} \frac{\p  Z_c^{-m}}{ \p g^2} Z_b^{-r/2}
\mathcal G_0^{n'} +   Z_A^{-n/2} Z_c^{-m} \frac{\p  Z_b^{-r/2} }{ \p
g^2}     \mathcal G_0^{n'}\,,\nonumber\\
\end{eqnarray}
where we have replaced $(n + 2m + r)$ with $n'$ as a shorthand.
Next, we have to calculate all the derivatives w.r.t.~$g^2$.
\begin{itemize}
\item \textbf{Calculation of $ \frac{\p g_0^2 }{ \p g^2}$}\\
In dimensional regularization, with $d=4-\varepsilon$, one can write
down
\begin{eqnarray} \label{glabel}
g_0^2 &=& \mu^\varepsilon Z_g^2 g^2\,.
\end{eqnarray}
Hence, if we derive this equation w.r.t.~$g^2$,
\begin{eqnarray}\label{ggnul}
\frac{\p g_0^2 }{ \p g^2} &=& \mu^\varepsilon \frac{\p Z_g^2}{ \p
g^2} g^2 +  \mu^\varepsilon Z_g^2 ~=~ g_0^2 \left(  \frac{\p \ln
Z_g^2}{ \p g^2}  + \frac{1}{g^2}\right) \,.
\end{eqnarray}
From the previous equation, we still have to determine $ \frac{\p
\ln Z_g^2}{ \p g^2} $, which can be extracted from equation
\eqref{glabel}. Deriving this equation w.r.t.~$\mu$ gives,
\begin{eqnarray}
\mu \frac{\p g_0^2}{\p \mu} &=& \varepsilon \mu^\varepsilon Z_g^2
g^2 + \mu^\varepsilon  \frac{\p Z_g^2}{\p g^2} \mu\frac{\p g^2}{\p
\mu} g^2 + \mu^\varepsilon Z_g^2 \mu \frac{\p g^2}{\p \mu} = 0\,,
\end{eqnarray}
were we have applied the chain rule again. We can rewrite this
equation making use of the definition of the $\beta$-funtion
\begin{eqnarray}
\mu\frac{\p g^2}{\p \mu} &=& - \varepsilon g^2 + \beta(g^2)\,,
\end{eqnarray}
where we have immediately extracted the part in $\varepsilon$, and
we obtain,
\begin{eqnarray}
   \frac{\p \ln Z_g^2}{\p g^2}   &=&  \frac{1}{g^2} \left( \frac{-\varepsilon  g^2 }{\mu\frac{\p g^2}{\p \mu}}-1 \right) ~=~ \frac{1}{g^2} \left( \frac{ -\beta(g^2)}{  - \varepsilon g^2 + \beta(g^2) }
   \right)\,.
\end{eqnarray}
If we insert this result into expression \eqref{ggnul}, we
ultimately find
\begin{eqnarray}
\frac{\p g_0^2 }{ \p g^2} &=& \frac{- \varepsilon g_0^2}{ - \epsilon
g^2 + \beta(g^2)}\,.
\end{eqnarray}

\item \textbf{Calculation of $ \frac{\p Z_A^{-n/2} }{ \p g^2} $}\\
The next derivative w.r.t.~$g^2$ can  be calculated in a
 similar way. We start by applying the chain rule,
\begin{eqnarray}\label{one}
\frac{\p  Z_A^{-n/2} }{ \p g^2} &=& - n \frac{ Z_A^{-n/2}
}{Z_A^{1/2}} \frac{\p Z_A^{1/2}}{ \p g^2} ~=~ - n Z_A^{-n/2} \frac{
\p \ln Z_A^{1/2}}{ \p g^2}\,.
\end{eqnarray}
Next, we derive $ \frac{ \p \ln Z_A^{1/2}}{ \p g^2}$ from the
definition of the gluon anomalous dimension,
\begin{eqnarray}\label{two}
\gamma_A &=&\mu \frac{\p \ln Z_A^{1/2}}{\p \mu}  ~=~ \mu \frac{\p
g^2 }{\p \mu}  \frac{\p \ln Z_A^{1/2}}{\p g^2} ~=~ \left(  -
\varepsilon g^2 + \beta(g^2) \right) \frac{\p \ln Z_A^{1/2}}{\p
g^2}\,.
\end{eqnarray}
From expression \eqref{one} and \eqref{two}, it now follows
\begin{eqnarray}
\frac{\p  Z_A^{-n/2} }{ \p g^2} &=& -n  Z_A^{-n/2} \frac{ \gamma_A}{
- \varepsilon g^2 + \beta(g^2)}\,.
\end{eqnarray}

\item \textbf{Calculation of $ \frac{\p Z_c^{-m} }{ \p g^2} $}\\
Completely analogously, we find with the help of the anomalous
dimension of the ghost field,
\begin{eqnarray}
 \mu \frac{\p Z_c^{1/2}}{\p \mu} &=& \gamma_c Z_c^{1/2}\,,
\end{eqnarray}
\begin{eqnarray}
\frac{\p  Z_c^{-m} }{ \p g^2} &=& -2m  Z_A^{-n/2} \frac{ \gamma_A}{
- \varepsilon g^2 + \beta(g^2)}\,.
\end{eqnarray}

\item \textbf{Calculation of $ \frac{\p Z_b^{-r/2} }{ \p g^2} $}\\
Finally, from
\begin{eqnarray}
 \mu \frac{\p Z_b^{1/2}}{\p \mu} &=& \gamma_b Z_c^{1/2}\,,
\end{eqnarray}
we deduce
\begin{eqnarray}
\frac{\p Z_b^{-r/2}  }{ \p g^2} &=& -r Z_A^{-n/2} \frac{ \gamma_A}{
- \varepsilon g^2 + \beta(g^2)}\,.
\end{eqnarray}
\end{itemize}
Taking all the previous results into account, we can rewrite
expression \eqref{finalgoal2},
\begin{eqnarray}\label{finalgoal3}
\frac{\d \mathcal G^{n'} }{\d g^2} &=& \frac{Z_A^{-n/2}Z_c^{-m}
Z_b^{-r/2} }{ - \varepsilon g^2 + \beta} \left[-\varepsilon g^2_0
\frac{\p}{\p g_0^2} - n \gamma_A -  2m \gamma_c -  r \gamma_b
\right] \mathcal G_0^{n'}\,.
\end{eqnarray}
The right hand side of \eqref{finalgoal3} still contains bare
quantities, which we have to rewrite in terms of renormalized
quantities. Notice also that we would like to get rid of the field
numbers $n$, $m$ and $r$ as the mixing matrix will evidently will be
independent of these numbers as they are arbitrary.

We shall now alter the right hand side of equation
\eqref{finalgoal3} by calculating $\frac{\p}{\p g_0^2} \mathcal
G_0^{n'}$ and by removing the fields numbers. After a little bit of
algebra, we obtain,
\begin{eqnarray}
\frac{\p (\e^{-S})}{\p g_0^2} &=&-\int \d^4 y \left( -
\frac{1}{g_0^2}  \left[ \frac{F_0^2(y)}{4}\right]  +
\frac{1}{2g_0^2} \left[ A_0(y)\frac{\delta S}{\delta A_0(y)}
\right]- \frac{1}{2g_0^2} \left[ b_0(y) \p A_0(y)\right] \right)
\e^{-S}\,.
\end{eqnarray}
Consequently, deriving the $n'$ point Green function $\mathcal
G^{n'}_0$ w.r.t.~$g^2_0$ will result in several insertions in this
Green function,
\begin{eqnarray}
g_0^2 \frac{d \mathcal G^{n'}_0}{ d g_0} &=&  \int \d^4 y \left(
\mathcal G^{n'}_0 \biggl\{ \frac{F_0^2(y)}{4} \biggr\} - \frac{1}{2}
\mathcal G^{n'}_0 \biggl\{ A_0(y)\frac{\delta S}{\delta A_0(y)}
\biggr\}  + \frac{1}{2} \mathcal G^{n'}_0 \biggl\{ b_0(y) \p A_0(y)
\biggr\} \right)\,,
\end{eqnarray}
where we have introduced a shorthand notation, e.g.
\begin{eqnarray}
\mathcal G^{n'}_0 \biggl\{ \frac{F_0^2(y)}{4} \biggr\} &=& \Braket{
\frac{F_0^2(y)}{4} A(x_1)\ldots A(x_n) c(y_1)\ldots c(y_m )
\overline c(\hat y_1) \overline c (\hat y_m) b(z_1)\ldots b(z_r)
}\,.
\end{eqnarray}
The field numbers can be rewritten by inserting the corresponding
counting operator. If we start by inserting the counting operator
for the gluon fields $n$, we find
\begin{eqnarray}
 \int \d^4 y \mathcal G_0^{n'}\biggl\{ A_0(y) \frac{\delta S}{\delta A_0(y)} \biggr\}  &=&n \mathcal
 G_0^{n'}\,,
\end{eqnarray}
as derived in equation \eqref{count}. We can derive analogous
relations for the other counting operators,
\begin{eqnarray}
 \int \d^4 y \mathcal G_0^{n'}\biggl\{  c_0(y) \frac{\delta S}{\delta  c_0(y)} \biggr\} &=&m \mathcal G_0^{n'}\,, \nonumber\\
 \int \d^4 y \mathcal G_0^{n'}\biggl\{ b_0(y) \frac{\delta S}{\delta b_0(y)} \biggr\}  &=&r \mathcal
 G_0^{n'}\,.
\end{eqnarray}
Taking all these results together, expression \eqref{finalgoal3} now
becomes,
\begin{eqnarray}
\frac{d \mathcal G^{n'} }{d g^2} &=& \frac{Z_A^{-n/2}Z_c^{-m} Z_b^{-r/2}}{-\varepsilon g^2 + \beta(g^2)} \int \d^4 y \Biggl[ - \epsilon \left(  \mathcal G_0^{n'}\left\{\mathcal F_0 (y)\right\} - \frac{1}{2} \mathcal G_0^{n'}\left\{\mathcal H_0 (y)  \right\} + \frac{1}{2} \mathcal G_0^{n'}\left\{\mathcal I_0 (y)  \right\} \right) \nonumber\\
 && -\gamma_A(g^2) \mathcal G_0^{n'}\left\{ \mathcal H_0(y) \right\} - 2  \gamma_c(g^2) \mathcal G_0^{n'}\left\{ \mathcal K_0(y) \right\}  -\gamma_b(g^2) \mathcal G_0^{n'}\left\{ \mathcal I_0(y)
 \right\}\Biggr]\,.
\end{eqnarray}
We have again introduced a notational shorthand for the equation of
motion operators, with $I_0 = b_0 \frac{\delta S}{\delta b_0}$ and
$\mbox{$K_0 = c_0 \frac{\delta S}{\delta c_0}$}$ and with $\mathcal
F$ and $\mathcal H$ already defined before.

In the last part of the manipulation of the $n'$-point Green
function we reexpress all the operators again in terms of their
renormalized counterparts, thereby writing all the divergences
explicitly in terms of $\varepsilon$. Firstly, we can reabsorb the
$Z$-factors into $\mathcal G_0^{n'}$ to find,
\begin{eqnarray}\label{finalgoal4}
\frac{d \mathcal G^{n'} }{d g^2} &=& \frac{1}{-\varepsilon g^2 + \beta(g^2)} \int \d^4 y \Biggl[ - \varepsilon \left(  \mathcal G^{n'}\left\{\mathcal F_0 (y)\right\} - \frac{1}{2} \mathcal G^{n'}\left\{\mathcal H_0 (y)  \right\} + \frac{1}{2} \mathcal G_0^{n'}\left\{\mathcal I_0 (y)  \right\} \right) \nonumber\\
  && -\gamma_A(g^2) \mathcal G^{n'}\left\{ \mathcal H_0(y) \right\} - 2  \gamma_c(g^2) \mathcal G^{n'}\left\{ \mathcal K_0(y) \right\}  -\gamma_b(g^2) \mathcal G^{n'}\left\{ \mathcal I_0(y)
  \right\}\Biggr]\,.
\end{eqnarray}
Secondly, we parametrize the mixing matrix \eqref{operatormatrix},
\begin{eqnarray}
 \left(
  \begin{array}{c}
    \mathcal F_0 \\
    \mathcal L_0 \\
    \mathcal H_0
  \end{array}
\right) &= & \left(
          \begin{array}{ccc}
            1 + \frac{a}{\varepsilon}& -\frac{b}{\varepsilon}  &-\frac{b}{\varepsilon} \\
            0  &1 & 0   \\
           0 & 0& 1          \end{array}
        \right)
        \left(
\begin{array}{c}
    \mathcal F \\
    \mathcal L \\
    \mathcal H
  \end{array}
\right)\,.
\end{eqnarray}
Here we have displayed the fact that the entries associated with
$a(g^2, \varepsilon)$ and $b(g^2, \varepsilon)$, which represent a
formal power series in $g^2$, must at least have a simple pole in
$\varepsilon$. We recall that the integrated operator $\mathcal L_0$
is proportional to the sum of the two counting operators
$\int\mathcal I_0$ and $\int \mathcal K_0$, see expression
\eqref{countcon}. Therefore $\int\mathcal I_0 = \int\mathcal I$ and
$\int\mathcal K_0 =\int\mathcal K$. Inserting all this information
into expression \eqref{finalgoal4} yields,
\begin{eqnarray}\label{finalgoal5}
\frac{\p \mathcal G^{n'} }{\p g^2} &=&   \frac{1}{-\varepsilon g^2 + \beta(g^2) }\int \d^4 y \Biggl[    \mathcal G^{n'}\Bigl\{ (- \varepsilon -a) \mathcal F  (y) - b \mathcal I(y)- b \mathcal K(y) +b \mathcal H(y)  + \frac{\varepsilon}{2} \mathcal H (y)  -\frac{\varepsilon}{2} \mathcal I (y) \nonumber\\
&& \hspace{5.5cm} -\gamma_A(g^2) \mathcal H(y) - 2  \gamma_c(g^2)
\mathcal K(y)   -\gamma_b(g^2)  \mathcal I(y)\Bigr\} \Biggr]\,,
\end{eqnarray}
giving us the final result from which we shall be able to fix the
matrix elements of expression \eqref{operatormatrix}.

As the left hand side of our final expression \eqref{finalgoal5} is
finite, the right hand side is finite too. Therefore, the following
coefficients must be finite,
\begin{eqnarray}
\mathcal F &:& \frac{- \varepsilon -a }{-\varepsilon g^2 + \beta(g^2) } = \frac{1}{g^2} \frac{(1  +a/ \varepsilon) }{1 - \beta(g^2) /(\varepsilon g^2) }\,, \nonumber\\
\mathcal I &:& \frac{ -\varepsilon/2 - b -  \gamma_b(g^2)  }{-\varepsilon g^2 + \beta(g^2) } = \frac{1}{2 g^2} \frac{ 1 + 2(b +  \gamma_b(g^2))/\varepsilon  }{1  - \beta(g^2)/(\varepsilon g^2) }\,,   \nonumber\\
\mathcal H &:& \frac{ \varepsilon/2 + b - \gamma_A(g^2)  }{-\varepsilon g^2 + \beta(g^2) } = -\frac{1}{2 g^2} \frac{1  + 2( b - \gamma_A(g^2))/\varepsilon }{1 - \beta(g^2) /(\varepsilon g^2) }\,, \nonumber\\
\mathcal K &:& \frac{  -b -  2\gamma_c(g^2)  }{-\varepsilon g^2 +
\beta(g^2) }\,,
\end{eqnarray}
seen as a power series in $g^2$. This can only be true if
\begin{eqnarray}
a(g^2,\varepsilon) &=& - \frac{\beta(g^2)}{g^2}\,, \nonumber\\
b(g^2,\varepsilon)&=& \gamma_A(g^2) -\frac{1}{2}\frac{
\beta(g^2)}{g^2} = -\gamma_b(g^2) -\frac{1}{2}\frac{
\beta(g^2)}{g^2} = -2\gamma_c(g^2)\,.
\end{eqnarray}
In fact, this last equation reveals a connection between the
anomalous dimension of $A$ and $b$, and between the anomalous
dimension of $A$, $g$ and $c$, namely
\begin{eqnarray}\label{rel}
\gamma_A + \gamma_b &=& 0\,, \nonumber\\
\gamma_A + 2 \gamma_c &=& \frac{ \beta }{ 2 g^2}\,.
 \end{eqnarray}
These relations are well-known to hold in the Landau gauge, since
$Z_A Z_b = 1$ and $Z_c Z^{1/2}_A Z_g = 1$, as derived from the
algebraic renormalization analysis, which  leads to equations
\eqref{Z1} and \eqref{Z2}.

In summary, we have completely fixed the mixing matrix in term of
the elementary renormalization group functions, and this to all
orders of perturbation theory,
\begin{eqnarray}
\underbrace{ \left(
  \begin{array}{c}
    \mathcal F_0 \\
    \mathcal L_0 \\
    \mathcal H_0
  \end{array}
\right)}_{X_0} &= & \underbrace{\left(
          \begin{array}{ccc}
            1 -\frac{\beta(g^2) / g^2}{\varepsilon}& -\frac{2 \gamma_c(g^2)}{\varepsilon}  &-\frac{ 2 \gamma_c(g^2)}{\varepsilon} \\
            0  &1 & 0   \\
           0 & 0& 1          \end{array}
        \right)}_{Z}
       \underbrace{ \left(
\begin{array}{c}
    \mathcal F \\
    \mathcal L \\
    \mathcal H
  \end{array}
\right)}_{X}\,.
\end{eqnarray}
In addition, as a check of this result, we have also uncovered two
relations, \eqref{rel}, between anomalous dimensions which must hold
for consistency. These correspond to \eqref{Z1} and \eqref{Z2},
which are well-known nonrenormalization theorems in the Landau
gauge.

\sect{Constructing a renormalization group invariant} As a last
step, we can now look for a renormalization group invariant operator
by determining the anomalous dimension $\Gamma$ coming from the
mixing matrix $Z$. We define the anomalous dimension matrix $\Gamma$
as
\begin{eqnarray}
\mu \frac{\p}{\p \mu} Z &=& Z\, \Gamma\,.
\end{eqnarray}
For the calculation of $\Gamma$, we require
\begin{eqnarray}
\mu \frac{\p}{\p \mu} \left(1 -\frac{\beta / g^2}{\varepsilon}\right) &=& \frac{1}{\epsilon} (- \varepsilon g^2 + \beta(g^2)) \frac{\p (\beta/g^2)}{\p g^2} \nonumber\\
- \mu \frac{\p}{\p \mu} \frac{ \gamma_c}{\varepsilon} &=&
\frac{1}\varepsilon{}(- \varepsilon g^2 + \beta(g^2)) \frac{\p
\gamma_c}{\p g^2}\,,
\end{eqnarray}
so we obtain,
\begin{eqnarray}
 \Gamma &= & \left(
          \begin{array}{ccc}
            g^2 \frac{\p (\beta / g^2)}{\p g^2}& -g^2 \frac{\p \gamma_c}{\p g^2}  &-g^2\frac{ \p \gamma_c}{\p g^2} \\
            0  &0 & 0   \\
           0 & 0& 0          \end{array}
        \right)\,,
\end{eqnarray}
which is indeed finite, a nice consistency check. This matrix is
then related to the anomalous dimension of the operators:
\begin{eqnarray}
X_0 ~=~ Z X
&\Rightarrow& 0 ~=~ \mu \frac{\p Z}{ \p \mu}  X + Z \mu \frac{\p X}{\p \mu} \nonumber\\
&\Rightarrow& \mu \frac{\p X}{\p \mu} ~=~ - \Gamma X \,.
\end{eqnarray}
We now have all the ingredients at our disposal to determine a
renormalization group invariant operator. We are looking for a
linear combination of $\mathcal F$, $\mathcal L$ and $\mathcal H$
which does not run,
\begin{eqnarray}
\mu \frac{\p}{\p \mu} \left[ k \mathcal F + \ell  \mathcal G + m
\mathcal H \right] &=& 0\,,
\end{eqnarray}
whereby $k$, $\ell$ and $m$ are to be understood as functions of
$g^2$. Invoking the chain rule gives
\begin{eqnarray}
\mu \frac{\p k}{\p \mu}  \mathcal F  - k  g^2 \frac{\p (\beta /
g^2)}{\p g^2} \mathcal F  + k g^2 \frac{\p \gamma_c}{\p g^2}
\mathcal L + k g^2 \frac{\p \gamma_c}{\p g^2} \mathcal H  + \mu
\frac{\p \ell }{\p \mu}  \mathcal L   + \mu \frac{\p m }{\p \mu}
\mathcal H  &=& 0\,.
\end{eqnarray}
This previous equation results in two differential equations,
\begin{eqnarray}\begin{cases}
\mu \frac{\p k}{\p \mu}   - k  g^2 \frac{\p (\beta / g^2)}{\p g^2}  =0\,, \nonumber\\
\mu \frac{\p \ell }{\p \mu} + k g^2 \frac{\p \gamma_c}{\p g^2} = 0\,, \nonumber\\
\ell = m\,, \end{cases}
\end{eqnarray}
which can be solved by,
\begin{eqnarray}\begin{cases}
k(g^2) = \frac{\beta(g^2)}{g^2}\,, \nonumber\\
\ell(g^2)  = m(g^2) = -\gamma_c(g^2)\,.\end{cases}
\end{eqnarray}
In summary, we have determined a renormalization group invariant
scalar operator $\mathcal{R}$ containing $\mathcal F$. Explicitly,
\begin{eqnarray}
\mathcal{R}&=&\frac{1}{4}\frac{\beta(g^2)}{g^2}F_{\mu\nu}^2
-\gamma_c(g^2)\left(A_\mu^a\p_\mu b^a  + \p_\mu \overline c^a
D_\mu^{ab} c^{b}\right)- \gamma_c(g^2) A_{\mu}^a\frac{\delta
S}{\delta A_\mu^a}\,,
\end{eqnarray}
without having calculated any loop diagram. Moreover, this invariant
is equal to the trace anomaly $\Theta_\mu^\mu$, which is expected as
$\Theta_\mu^\mu$ is also a $d=4$ renormalization group invariant
\cite{Collins:1976yq}.

\sect{Conclusion} In this paper, we have provided a detailed
analysis of the renormalization of the non-integrated operator
$F_{\mu\nu}^2$ in Yang-Mills gauge theory in the Landau gauge, and
this to all orders. We have shown that this operator mixes with two
other operators: the BRST exact operator $s(\overline c^a \p_\mu
A_\mu^a)$ and a gluon equation of motion operator, $A_\mu^a
\frac{\delta S}{\delta A_\mu^a}$. We have composed the corresponding
renormalization matrix $Z$ and we have been able to determine this
matrix to all orders in perturbation theory. Several checks have
confirmed the results. We have recovered the well known
non-renormalization theorems in the Landau gauge, relating the
gluon, ghost, auxiliary field and coupling constant anomalous
dimension. In addition, we have calculated the anomalous dimension
matrix using $Z$, which was nicely finite, and we have been able to
construct a renormalization group invariant containing
$F_{\mu\nu}^2$.

When turning to physical states, the BRST exact term and the
equation of motion term will drop and be no longer relevant.
However, we have paved the way for more complicated actions as this
framework is very solid. For example, in the Gribov-Zwanziger
action, when investigating the operator $F_{\mu\nu}^2$, we also
expect a mixing. Again, we would expect three different classes of
operators, the gauge invariant, the BRST exact and the ones which
vanish upon using the equations of motion. However, the
Gribov-Zwanziger action is no longer BRST invariant
\cite{Dudal:2008sp,Zwanziger:1992qr}. Fortunately, the framework we
have set up can be saved as one can embed the Gribov-Zwanziger
action into a ``larger'' BRST invariant action which will reduce to
the known Gribov-Zwanziger action in the so-called physical limit
\cite{Dudal:2008sp}. The embedding into a BRST
invariant action seems to be the crucial tool to correctly identify
the relevant, renormalizable, gauge invariant operators in the
Gribov-Zwanziger case, as the physical limit of their more general
counterparts constructed from the larger action. In this case, the
BRST exact operators shall no longer be automatically irrelevant
when taking the physical limit, as the BRST symmetry is softly
broken by the restriction to the Gribov region. Therefore, we expect
a non-trivial mixing to occur when we transfer to the physical
reality of glueballs, i.e. when calculating the corresponding
correlator. Hence, it will certainly be of the utmost importance to
take all the possible mixings into account in order to be able to
construct a renormalization group invariant \cite{nele}.

We did not include fermions in our analysis. The cohomological
treatment of massless fermions is not that complicated, see
\cite{Piguet:1995er}. However, additional equation of motion terms,
as well as the derivative of the singlet vector current can be
relevant. In addition, including non-degenerate fermion masses would
complicate matters, due to the introduction of a fermion mass mixing
matrix. As we already mentioned, our interest is and will be mainly
focused on pure gauge theories \cite{nele}.

\section*{Acknowledgments.}
We wish to thank J.~A.~Gracey for many helpful comments. D.~Dudal
and N.~Vandersickel are supported by the Research Foundation -
Flanders (FWO). The Conselho Nacional de Desenvolvimento
Cient\'{i}fico e Tecnol\'{o}gico (CNPq-Brazil), the Faperj,
Funda{\c{c}}{\~{a}}o de Amparo {\`{a}} Pesquisa do Estado do Rio de
Janeiro, the SR2-UERJ and the Coordena{\c{c}}{\~{a}}o de
Aperfei{\c{c}}oamento de Pessoal de N{\'{i}}vel Superior (CAPES) are
gratefully acknowledged for financial support. This work is
supported in part by funds provided by the US Department of Energy
(DOE) under cooperative research agreement DEFG02-05ER41360.
N.~Vandersickel acknowledges the hospitality at the CTP (MIT) and
UERJ, and D.~Dudal that at the UERJ, where parts of this work were
done.

\end{document}